\documentclass[twocolumn]{aastex631}
\graphicspath{{./}{../pictures/}}
\received{}
\revised{}
\accepted{}

\shorttitle{PopIII Stars Spectra}
\shortauthors{Hassan et al.}

\usepackage{graphicx}	
\usepackage{amsmath}	
\usepackage{amssymb}	
\usepackage{cases}
\usepackage{array,multirow}
\usepackage{upgreek}
\usepackage{textcomp}
\usepackage{mhchem}
\usepackage{longtable}

\usepackage{hyperref}
\usepackage[referable]{threeparttablex}
\usepackage{booktabs, dcolumn}

\def\apjl{ApJL}

\def\aap{A\&A}

\def\apjs{ApJ Supp.}

\usepackage[normalem]{ulem}

\begin{document}

\title{Spectral Evolution of Rotating Population III Stars}

\correspondingauthor{Jake B. Hassan}
\email{jakebhassan@gmail.com}

\author[0009-0004-5685-6155]{Jake B. Hassan}
\affil{Department of Physics and Astronomy, Stony Brook University, Stony Brook, NY 11794-3800, USA}
\author[0000-0002-3635-5677]{Rosalba Perna}
\affil{Department of Physics and Astronomy, Stony Brook University, Stony Brook, NY 11794-3800, USA}

\author[0000-0002-8171-8596]{Matteo Cantiello}
\affiliation{Center for Computational Astrophysics, Flatiron Institute, New York, NY 10010, USA}
\affiliation{Department of Astrophysical Sciences, Princeton University, Princeton, NJ 08544, USA}

\author[0000-0002-4299-2517]{Tyler M. Parsotan}
\affil{Astrophysics Science Division, NASA Goddard Space Flight Center, Greenbelt, MD 20771, USA}
\author[0000-0002-9190-662X]{Davide Lazzati}
\affil{Department of Physics, Oregon State University, 301 Weniger Hall, Corvallis, OR 97331, USA}
\author[0009-0008-1678-2787]{Nathan Walker}
\affil{Department of Physics, Oregon State University, 301 Weniger Hall, Corvallis, OR 97331, USA}

\begin{abstract}

Population III (Pop III) stars, the first generation of stars formed from primordial gas, played a fundamental role in shaping the early universe through their influence on cosmic reionization, early chemical enrichment, and the formation of the first galaxies. However, to date they have eluded direct detection due to their short lifetimes and high redshifts.  The launch of the James Webb Space Telescope (JWST) has revolutionized observational capabilities, providing the opportunity to detect Pop~III stars via caustic lensing, where strong gravitational lensing magnifies individual stars to observable levels. This prospect makes it compelling to develop accurate models for their spectral characteristics to distinguish them from other stellar populations. Previous studies have focused on computing the spectral properties of non-rotating, zero-age main sequence (ZAMS) Pop III stars. In this work, we expand upon these efforts by incorporating the effects of stellar rotation and post-ZAMS evolution into spectral calculations. We use the JWST bands and magnitude limits to identify the optimal observing conditions, both for isolated stars, as well as for small star clusters.
We find that, while rotation does not appreciably change the observability at ZAMS, the subsequent evolution can significantly brighten the stars, making the most massive ones potentially visible with only moderate lensing. 
\end{abstract}


\section{Introduction} 
\label{sec:intro}

Population~III stars (Pop~III), the universe's first generation of stars, formed
from primordial gas composed almost entirely of hydrogen and
helium (e.g. \citealt{Bromm2002,Abel2002,Turk2009}).  They 
 are theorized to form through multiple channels including self-gravity of neutral hydrogen with a primordial accretion disk surrounding the
proto-Pop III star \citep{stahler1986a,stahler1986b}, the fragmentation of the primordial gas on large scales \citep{clark2011}, and
the irradiation and cooling of primordial gas allowing clumps of gas to gravitate to one
another \citep{maio2011}.
These different formation channels affect the initial mass function of
Pop~III stars, which characterizes the initial distribution of stellar masses after formation.
 Nonetheless, various models predict that the initial mass
of Pop III stars can range between $\sim (10^{-1}-10^3)\;\mathrm{M}_\odot$ (e.g. \citealt{hosokawa2011,stacy2012,hirano2014,clark2011,greif2011,susa2014,reinoso2025}),
where the large masses are possible due
to the suppression of significant mass loss resulting from the low metallicity of these
stars (e.g. \citealt{wollenberg2020,jaura2022}).
Pop~III stars have been predicted to first appear around 
a redshift $z\approx 30$, peak in the redshift range 
$17\gtrsim z \gtrsim 10$, and then decline but still continue at least until $z\sim 6$ (e.g. \citealt{trenti2009,muratov2013}), and possibly even  down to $z\sim 3$ \citep{liu2020}.

Pop~III stars are believed to have played an important role in
the early universe by initiating the synthesis of heavier
elements and influencing the formation and evolution
of subsequent stellar populations.
Despite their significance, direct detection  has
remained elusive to date due to their short lifetimes and the very large
cosmological distances at which they are expected.

The launch of the James Webb Space Telescope (JWST) has produced a renewed interest in these objects.  
Although direct detection remains beyond reach—even at the lowest anticipated redshifts and with extremely long exposure times—a promising avenue involves the detection of a gravitationally lensed star \citep{rydberg2013, zackrisson2015, Windhorst2018}. This is particularly feasible during a caustic transit event, 
that is a gravitational lensing event where the source lies behind a caustic line:
magnification factors as high as $\mu\sim 10^7$ may be achieved in idealized cases of galaxy cluster lensing \citep{miraldaescude1991}, albeit such magnifications can be significantly diminished due to microlensing effects from stars within the cluster \citep{diego2019}.
Estimates by \citet{Windhorst2018} using blackbody spectra and magnifications of $\mu \sim 10^4-10^5$ suggest that a Pop~III detection may be likely by observations of $\sim 30$ lensing clusters for a decade.

Building upon these theoretical and observational frameworks, \citet{Larkin2023} coupled detailed stellar atmospheres and evolutionary star modelling to characterize the observable properties of zero-age main-sequence (ZAMS) Pop~III stars.
Their investigation encompassed a broad spectrum of stellar masses, generating synthetic photometry and theoretical color-magnitude diagrams tailored to JWST's Near-Infrared Camera (NIRCam) bands. They found that detection of the most massive star in their study, $\sim 800 \;\mathrm{M}_\odot$, requires a magnification $\mu \sim $~a~few~$\times 10^3$ at the higher redshift $z=17$, and of $\sim 10^3$ at the lowest redshift $z= 3$. Less massive stars require more stringent requirements on the lensing amplification, with a $\sim 100 \;\mathrm{M}_\odot$ star at $z=3$ needing $\mu\sim 10^4$.

The study of \citet{Larkin2023} has been fundamental in exploring the observational characteristics of non-rotating Pop~III stars during their ZAMS phase. Previous work, however, had shown that rotation and magnetic fields have an effect on stellar structure and chemical mixing. In particular, \citet{Yoon2012} presented a comprehensive grid of massive Pop~III star models, revealing that rotation significantly influences the evolution and final fate of these stars, with rapid rotators potentially undergoing chemically homogeneous evolution. They also noted that these fast rotators could lead to progenitors for high-redshift gamma-ray bursts (GRBs, e.g. \citealt{Yoon2005,Nagakura2012}).

In this work we extend and generalize the study by \citet{Larkin2023}, continuing the theoretical efforts needed to fully characterize the observable properties of Pop~III stars in the JWST era. Specifically, 
using the stellar evolutionary code \texttt{MESA} \citep{paxton2011,paxton2013,paxton2015,paxton2018,paxton2019} in combination with the spectral code \texttt{ATLAS} \citep{kurucz1970},
we explore the impact of stellar rotation on spectral properties and examine the effects of evolution across the main sequence. As mentioned above, rotation is particularly important as it plays a crucial role in the production of GRBs from some of these stars, and numerical simulations of Pop~III stars 
indicate high rotation rates, with surface velocities from about $50\%$ to nearly $100\%$ of the Keplerian rotational velocity \citep{stacy2013}.
Meanwhile, stars generally become more luminous over the course of main sequence evolution, enhancing detectability. The later stages of the main sequence are also characterized by redder colors, which alter the stars' positions in the color-magnitude diagram, influencing their potential observability with JWST.

Our paper is organized as follows: numerical methods are detailed in Section~\ref{sec:methods}.
Our results on the spectra and observability of individual Pop~III stars are presented in Section \ref{sec:results}, followed by an illustrative case of a star cluster in Section \ref{sec:clusters}; in both cases, we explore implications for observability in the JWST bands. In Section \ref{sec:GRB}, we discuss how the models developed in this study relate to the potential end states of Pop~III stars as gamma-ray burst progenitors.
Lastly, we summarize and conclude in Section~\ref{sec:summary}.

\section{Numerical Methods}\label{sec:methods}
\subsection{Star Evolution with MESA}
\begin{figure*}[ht!]
\centering
\includegraphics[width=\textwidth]{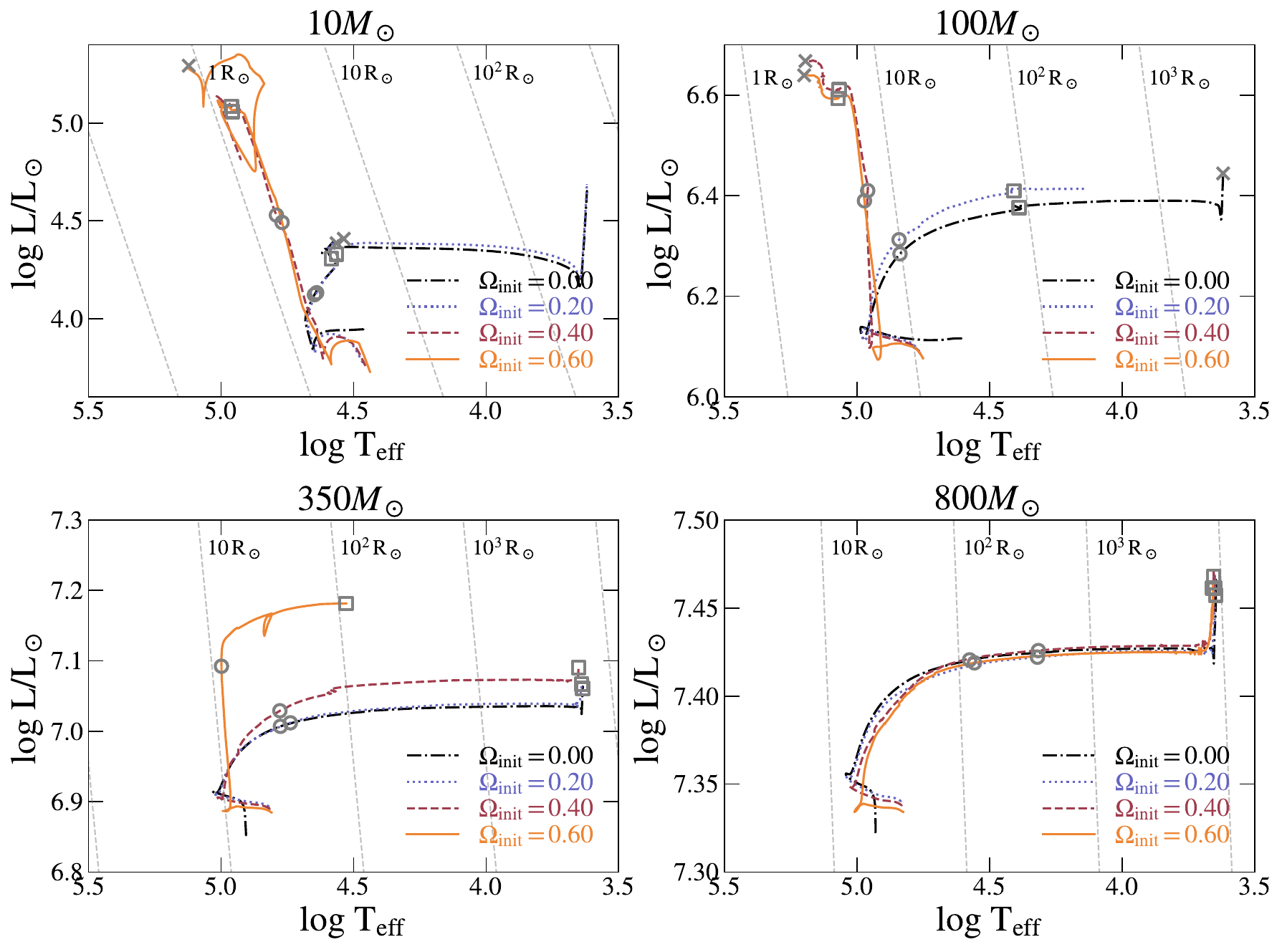}
\caption{HR-diagrams showing evolutionary tracks from the beginning of hydrogen burning to the end of the calculation for models of Pop III stars with different initial masses and rotational velocities. Circle symbols mark the position at which $X = 0.3$, while the end points of hydrogen burning and helium burning are marked with a square and cross, respectively.}
\label{fig:HRD}
\end{figure*}

\begin{figure*}[ht!]
\centering
\includegraphics[width=\textwidth]{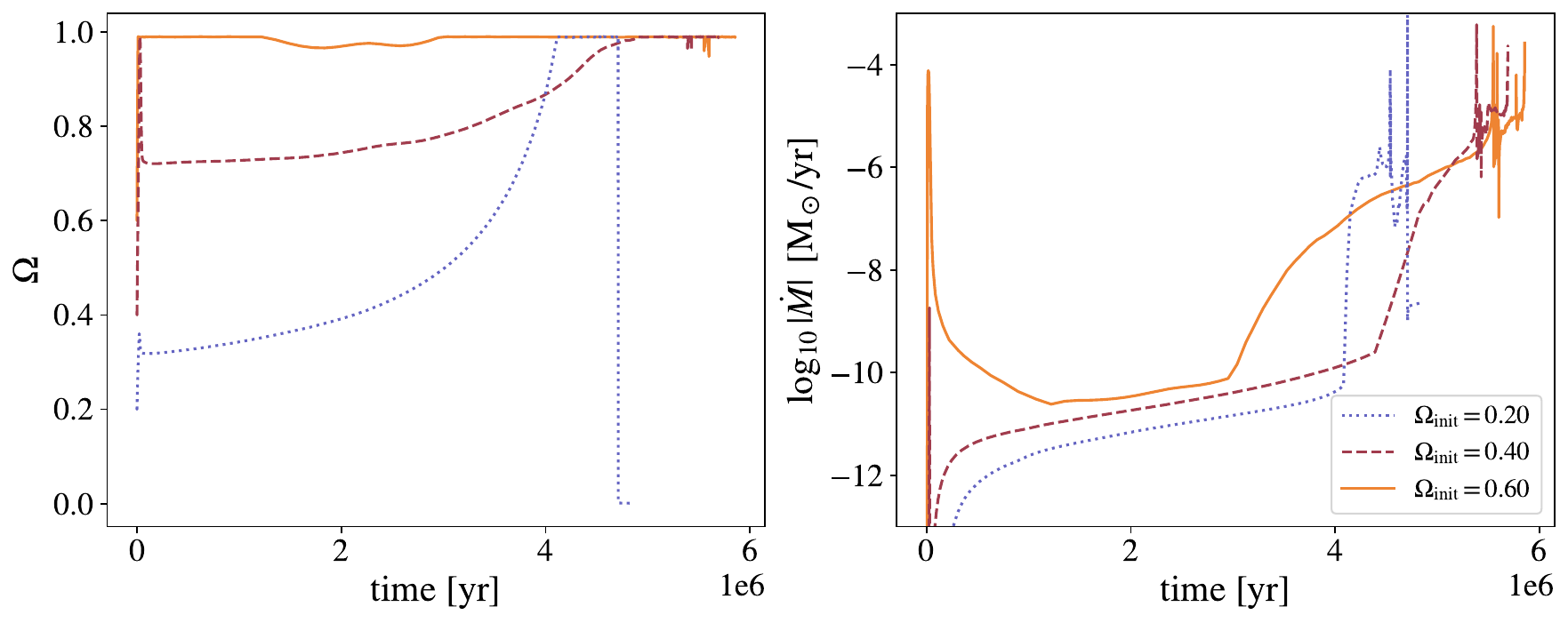}
\caption{Change in surface rotational velocity in units of the critical velocity (left panel) and mass loss rate (right panel) from pre-MS to helium exhaustion, for a 50 $\mathrm{M}_\odot$ model with varying initial velocities. Lines of constant radius are shown in gray.}
\label{fig:velocity}
\end{figure*}

We compute stellar evolutionary models of Pop~III stars using Modules for Experiments in Stellar Evolution (\texttt{MESA}, \citealt{paxton2011,paxton2013,paxton2015,paxton2018,paxton2019}), version r24.08.1. \texttt{MESA} is an open-source, one-dimensional stellar evolution code used for simulating the structure and evolution of stars across a wide range of masses, metallicities, and evolutionary phases. It solves the 
fully coupled equations of stellar structure using an implicit, adaptive timestep scheme. It includes several tabulated equations of state, covering a wide range of densities, temperatures, and compositions.

A total of 64 models were computed from the pre-main sequence (PMS) to core hydrogen depletion or beyond, across a grid of four initial rotational velocities ($v_{\mathrm{s}}/v_{\mathrm{crit}} = 0.0,\, 0.2,\, 0.4,\, 0.6$) and sixteen initial masses ranging from $M_\mathrm{init} = 10-800\;\mathrm{M}_\odot$. Since stars only spend a very small fraction of their life 
in post-MS stages as a whole,
the chance of observing Pop III stars in these later evolutionary phases is negligible. For this reason, we do not compute spectra for helium-burning phases and beyond.

In modeling the evolution of Pop III stars, our assumptions are largely based on those used in \cite{Yoon2012}. Models are initialized during the pre-main sequence with $Z=0$, $X=0.76$, $Y=0.24$ (\citealt{Yoon2012}) and a core temperature of $T_c = 5 \times 10^5 \;\mathrm{K}$ (\citealt{Larkin2023}). We use the Skye equation of state \citep{jermyn2021}.

Massive metal-free stars ($\gtrsim 20 \;\mathrm{M}_\odot$) cannot be supported by the proton-proton chain alone, and they initially do not have enough carbon to begin the CNO cycle. They therefore contract during the protostar phase until the triple-alpha process begins; once enough carbon is produced in the core ($X(\ce{C}) \sim 10^{-10}$), the CNO cycle begins and the star expands until it reaches hydrostatic equilibrium. To prevent this contraction from occurring until the star has finished relaxing toward the chosen parameters, we include a small amount of initial helium-3 ($X(\ce{^{3}He}) = 10^{-5}$). This allows the star to support itself via deuterium fusion during the relaxation period, after which the He-3 is quickly exhausted.

We define the zero-age main sequence (ZAMS) of our models as the point at which 90 percent of the star's luminosity comes from hydrogen burning; intermediate-age main sequence (IAMS) as the point at which $X_\mathrm{core} = 0.3$; and terminal-age main sequence (TAMS) is defined as the point at which $X_\mathrm{core} = 10^{-6}$.

We use the Ledoux criterion to determine the convective boundaries, using a mixing length parameter $\alpha_{\mathrm{MLT}} = 1.5$ and a semiconvection efficiency $\alpha_{\mathrm{SEM}} = 1.0$. We adopt the overshooting scheme for hydrogen-burning cores used in \cite{Yoon2012}.
For mass loss, we use the wind scheme described in \cite{Yoon2012}, neglecting wind entirely for nonrotating models. This is motivated by the fact that mass loss from stellar winds in metal-free, hot massive stars is found to be negligible \citep{Krticka2006}.
Far more dominant is rotationally-induced mass loss, occurring when the star's rotational velocity exceeds the critical point that can be supported by gravity \citep{paxton2013,perna2014}.

Figure~\ref{fig:HRD} shows Hertzsprung-Russell (HR) diagrams for four different initial masses, rotating at varying fractions of the critical velocity. 
Along the equator of a star of mass $M$ and radius $R$, this is defined as (\citealt{Yoon2012}):
\begin{equation}
v_{\text{crit}} = \sqrt{ \left(1 - \frac{L}{L_{\text{Edd}}} \right) \frac{GM}{R} }\,,
\end{equation}
where $L_{\rm Edd}=4\pi c\,GM/\kappa$, with $\kappa$ being the opacity. The ratio of the star's surface velocity $v_{\rm s}$ to the critical velocity is denoted $\Omega = v_{\rm s}/v_{\text{crit}}$. Note that \citet{Yoon2012} expresses rotation in terms of the Keplerian velocity $v_K = \sqrt{\frac{GM}{R}}$ rather than $v_\text{crit}$, which must be accounted for when comparing their results with ours.

For a given mass, the initial rotational speed (and hence the degree of chemical mixing) determines the evolutionary track of the star. 
Slowly-rotating models evolve redward, following canonical stellar evolution and becoming red supergiants (RSGs).
At higher rotational velocities, 
internal mixing processes such as the Eddington-Sweet circulation (e.g. \citealt{Maeder1987,Yoon2005,Woosley2006}) can become very efficient and prevent nuclear-burning from establishing a
compositional gradient, resulting in chemically homogeneous evolution (CHE). In CHE, models progress blueward as Wolf-Rayet stars, becoming much more compact and luminous during the main sequence than their RSG counterparts. This is because mixing results in more massive stellar cores and a larger mean molecular weight in the envelope. 

Models can follow a CHE initially, but then transition to canonical stellar evolution if they lose sufficient angular momentum via stellar winds. These transition models evolve blueward first, after which they expand and cool as blue supergiants (BSGs). Unlike the fast-rotating models which stay fully-chemically homogeneous well after the main sequence, BSG models manage to develop a compositional gradient in the envelope, thus regaining a core-envelope structure.

Throughout this work, our definitions for RSG, BSG and CHE follow \citet{Yoon2012}, who used the surface helium abundance at the end of core hydrogen burning ($Y_{s,\mathrm{TAMS}}$) as a criterion: models for which $Y_{s,\mathrm{TAMS}} \leq 0.7$ are considered to be on the RSG track; models for which $0.7< Y_{s,\mathrm{TAMS}} \leq 0.8$ are considered to be on the BSG track; and models for which $Y_{s,\mathrm{TAMS}} > 0.8$ are considered to be on the CHE track.

The most massive non-rotating models ($M_\mathrm{init}\gtrsim 150 \; \mathrm{M}_\odot$) develop very large convective cores while maintaining a core-envelope structure. This pushes their evolution close to the Hayashi track near the end of hydrogen burning. As a consequence, the location of the TAMS for these models occurs at surface temperatures similar to the location of helium exhaustion. 

As the initial mass of models below $\sim 150 \; \mathrm{M}_\odot$ is increased, CHE is generally reached at lower initial velocities. This is due to the increasing support of radiation in the envelope of massive stars helping the development of rotational instabilities \citep{Heger2000}. However, for our models above $\sim 200\; \mathrm{M}_\odot$, rotationally-induced mass loss and its associated angular momentum loss prevents CHE from occurring (see section 5 of \citealt{Yoon2012}). Some of these models can still become BSG at high initial rotational velocities, causing them to have high temperatures and luminosities during the main sequence before expanding redward during helium burning. Note that, compared to more slowly-rotating models, their luminosities on the red supergiant can still be much larger.

Figure~\ref{fig:velocity} shows the surface rotational velocity and mass loss rate as a function of time, for models with an initial mass of $50 \; \mathrm{M}_\odot$ and different initial rotational velocities. These models exhibit an increase in rotation during their contraction as protostars; for high initial velocities, the star may undergo critical rotation during this phase and shed some of its mass. Once the CNO cycle begins, the star expands and enters the main sequence, ending the initial increase in rotational velocity.

\subsection{Spectral Models for the Atmosphere}
Synthetic spectroscopy was created for each \texttt{MESA} model using \texttt{BasicATLAS} \citep{Gerasimov2022, Larkin2023}, a \texttt{Python} wrapper for \texttt{ATLAS 9} (\citealt{kurucz1970,kurucz2014}). This is an atmosphere modelling program written in \texttt{FORTRAN}
and uses Opacity Distribution Functions (ODFs) for treating line opacity. It 
calculates temperature and pressure profiles based on the effective temperature, composition and surface gravity in the atmosphere. The \texttt{ATLAS 9} suite also includes the \texttt{SYNTHE} program
\citep{kurucz1981}, which generates a spectrum based on the atmospheric profile, as well as the \texttt{DFSYNTHE}
program \citep{Castelli2005}, which generates opacity tables for a given atmospheric composition.

\texttt{ATLAS 9} takes the surface metallicity $\mathrm{[M/H]}$ and enhancements of individual metals $\mathrm{[E/M]}$ as input parameters. Since \texttt{MESA} reports the star's composition in terms of mass fractions, we obtained these parameters by first calculating the surface number density of each element:
\begin{equation}
    n_\mathrm{E}= \rho N_A  \frac{X(\mathrm{E})}{A_\mathrm{E}}\,,
\end{equation}
where $N_A$ is the Avogadro constant, $\rho$ is the baryon mass density, and $X(\mathrm{E})$ and $A_\mathrm{E}$ are the mass fraction and atomic mass of element $\mathrm{E}$, respectively. We define the abundance of the element as
\begin{equation}
    \epsilon_\mathrm{E}=12.00 + \log_{10} \left (\frac{n_\mathrm{E}}{n_\mathrm{H}} \right )\,.
\end{equation}
The abundance ratio for the element is then obtained via $\mathrm{[E/H]} = \epsilon_{\mathrm{E},\star}-\epsilon_{\mathrm{E},\odot}$, where $\star$ and $\odot$ denote the abundances in our star and in the Sun, respectively.\footnote{For a list of standard solar abundances, see \url{https://atmos.ucsd.edu/?p=solar.}} Using the overall metal number density $n_\mathrm{M}=\sum_{\mathrm{metals \; E}} n_\mathrm{E}$, we likewise obtain $\mathrm{[M/H]}$, one of the input parameters for \texttt{ATLAS 9}. The remaining input parameters are the enhancements of each metal, given by
\begin{equation}
    \mathrm{\left [\frac{E}{M} \right ]} = \mathrm{\left [\frac{E}{H} \right ]} - \mathrm{\left [\frac{M}{H} \right ]}\,.
\end{equation}

Our ODFs were created with \texttt{DFSYNTHE} and sampled at 57 temperatures. By default, these temperatures range from $10^{3.3} - 10^{5.3} \; \mathrm{K}$; while this is sufficient for lower mass stars, stellar atmospheres can reach far greater temperatures than this in more massive stars. Following \citet{Larkin2023}, we instead use the temperature range $10^{3.52} - 10^{5.85} \; \mathrm{K}$ in cases where the base of the star's atmosphere ($\tau = 100$) exceeds $10^{5.3} \; \mathrm{K}$.

As in \citet{Larkin2023}, the same two sets of ODFs (corresponding to the two aforementioned temperature ranges) can be used for all nonrotating models or models at ZAMS. For rotating models beyond ZAMS, rotational mixing can dredge core helium and metals up to the atmosphere, requiring us to calculate a unique ODF for each model. This increase in the surface helium abundance becomes especially significant for CHE stars near the end of main sequence.

Synthetic spectra were created with \texttt{SYNTHE}. We use the same criteria as \citet{Larkin2023} to determine whether the default wavelength range of $\sim$ 9nm $-$ 160 µm is sufficient for our model. In cases where it is not, we extend the range down to 4nm; if this is still not sufficient, we extend it to 0.1nm.

\section{Results}\label{sec:results}
\subsection{Spectral dependence on rotation and evolution}
\begin{figure*}[ht!]
\centering
\includegraphics[width=\textwidth]{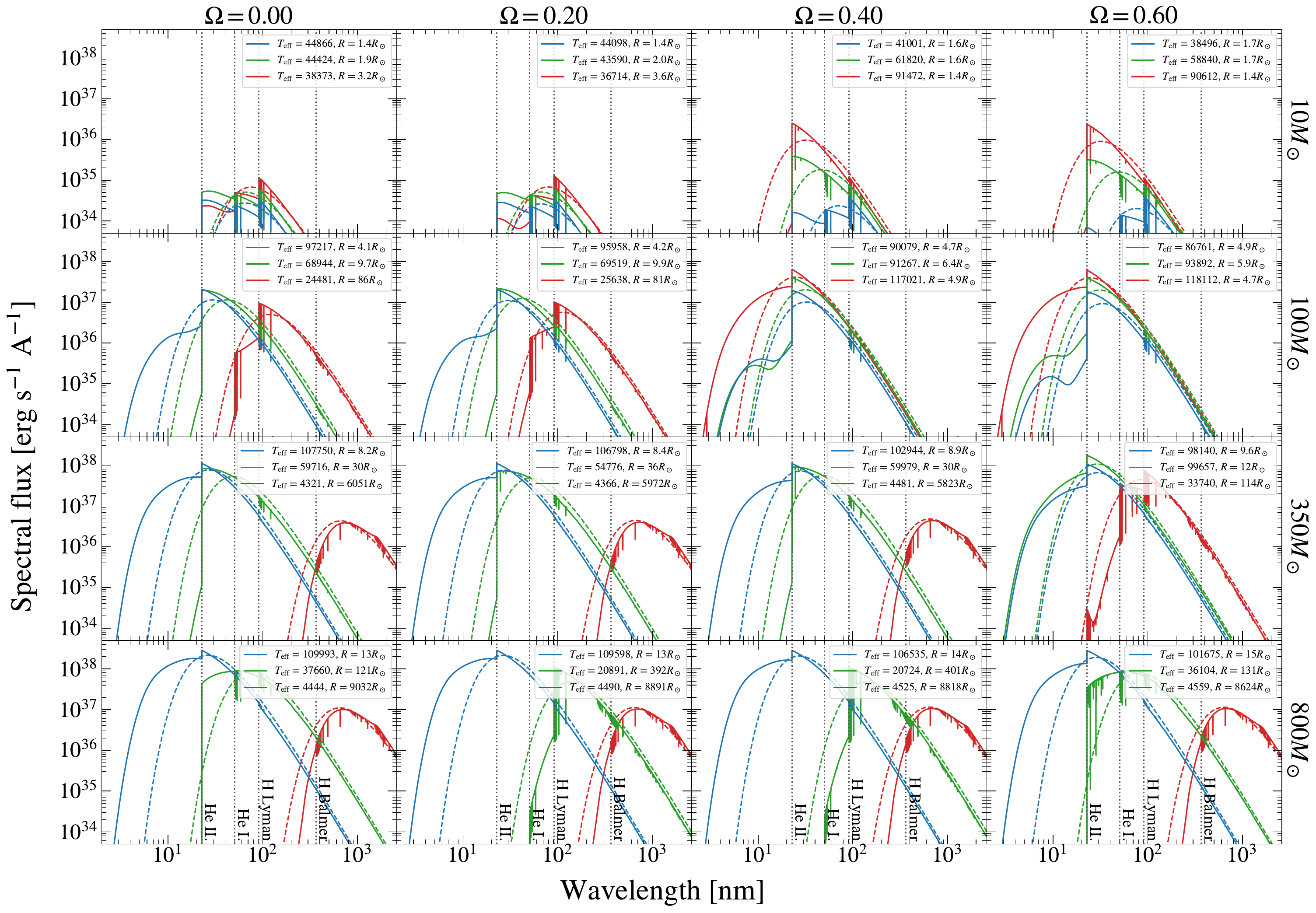}
\caption{Comparison of spectral fluxes and evolutionary tracks for Pop III stars with different initial masses and rotational velocities. Spectral fluxes are shown at ZAMS (blue), IAMS (green), and TAMS (red), which are the points indicated in the H-R diagram of Fig.~\ref{fig:HRD}. The corresponding blackbody models at the same effective temperatures and radii are shown with dashed curves. Photoionization breaks are indicated with gray dashed lines. }
\label{fig:spectra}
\end{figure*}

Figure \ref{fig:spectra} shows spectral fluxes corresponding to the same models as in Figure \ref{fig:HRD}, at different points on the main sequence. As expected, the spectra of CHE models peak at lower wavelengths. At later stages, the increasing helium surface abundance generally results in noticeable spikes associated with helium bound-bound interactions, as well as a significant drop in the spectrum beyond the helium photoionization line. Conversely, the decreasing surface hydrogen abundance reduces absorption at wavelengths associated with hydrogen ionization, but not helium.

\begin{figure*}[ht!]
\centering
\includegraphics[width=\textwidth]{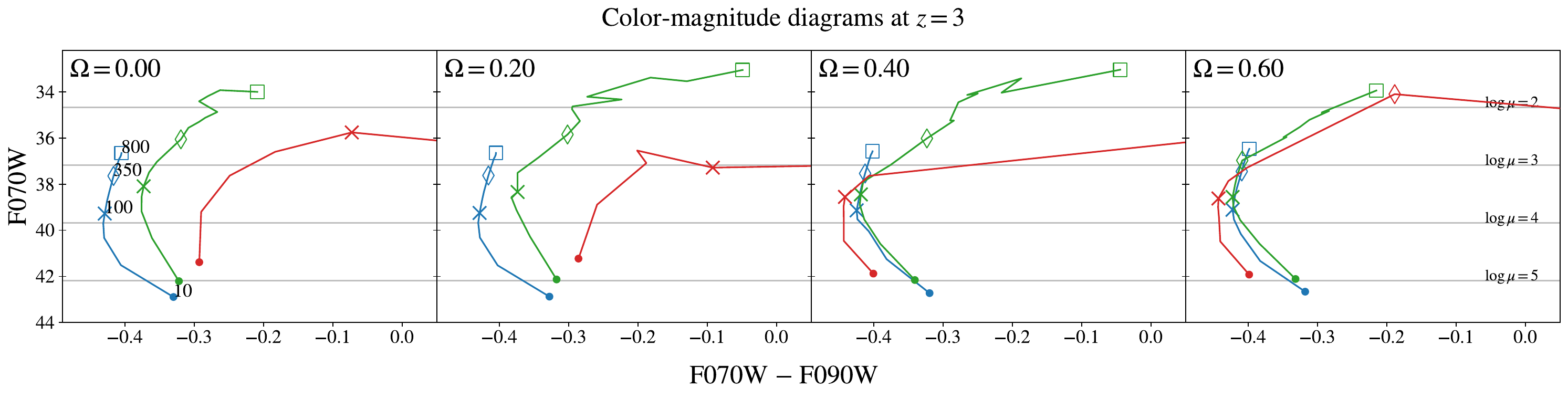}
\includegraphics[width=\textwidth]{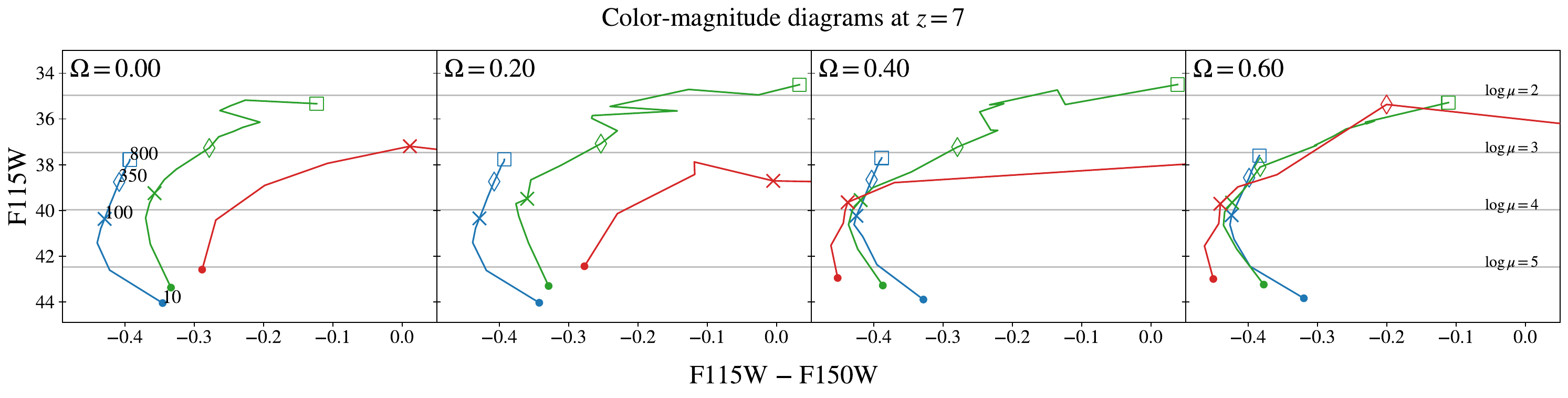}
\includegraphics[width=\textwidth]{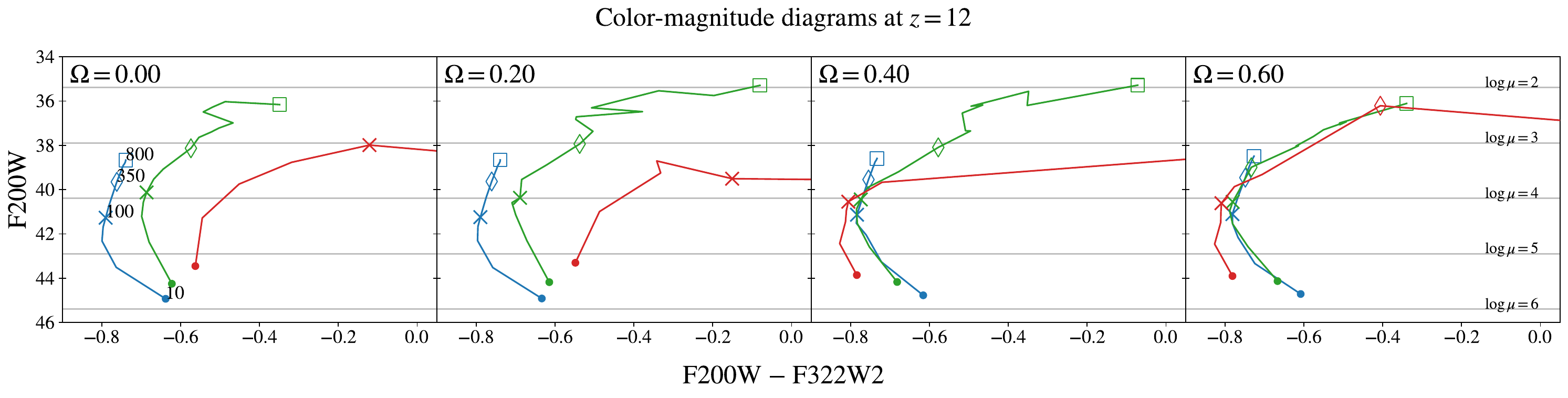}
\includegraphics[width=\textwidth]{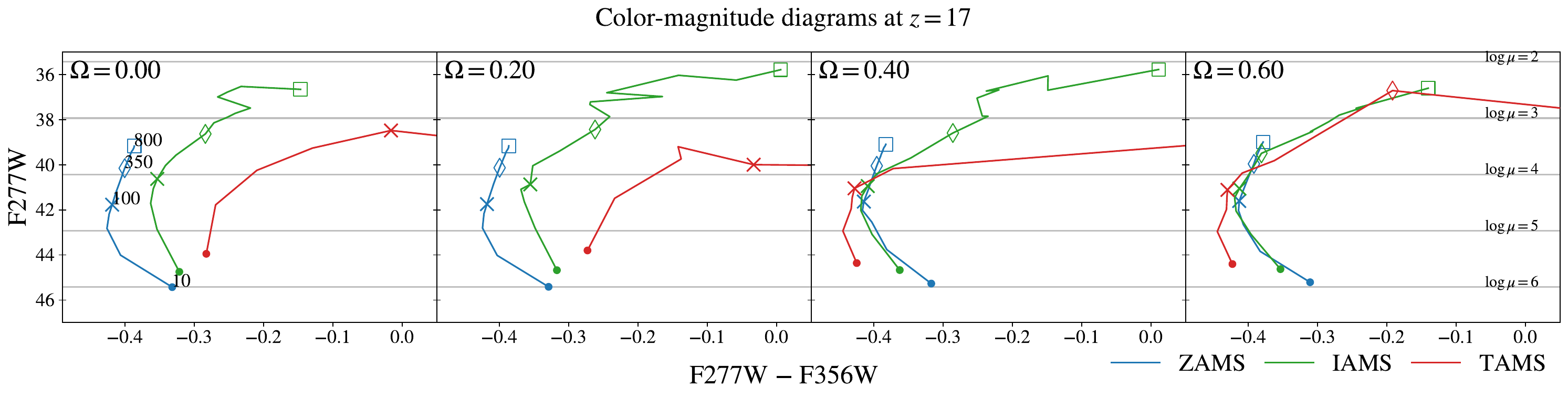}
\caption{Color-magnitude diagrams at different redshifts and rotational velocities. Results are shown for ZAMS (blue), IAMS (green) and TAMS (red). The models from Figure \ref{fig:HRD} are marked with unique symbols corresponding to their initial mass, which are labelled (in solar masses) on the ZAMS curve of the leftmost plots. The bands used are generally the optimal pair for all models at a given redshift, with the exception of massive RSGs at TAMS. Observability limits under various magnifications $\mu$ are shown with gray lines and labelled on the rightmost plots.}
\label{fig:colormagnitude}
\end{figure*}

\subsection{Detectability with JWST}
In the following, we use the synthetic spectra 
presented in the previous section to
compute magnitude and color estimates for the relevant JWST observation bands. 
We use the \texttt{ABMAG} magnitude system \citep{Oke1983}, which defines the relative magnitude $m_{\rm AB}$ as 
\begin{equation}
m_{\rm AB} = -2.5\,\log_{10} 
\left(\frac{\int \lambda \,f_\lambda(\lambda)\,e(\lambda) \,d\lambda}{\int 3631\,{\rm Jy} \,(c/\lambda)\, e(\lambda)\, d\lambda }\right)\,,
    \label{eq:mab}
\end{equation}
where $e(\lambda)$ is the instrumental efficiency and $f_\lambda(\lambda)$ is the flux density at the observer.
For a star of radius $R$, this is 
related to the
emitted flux density $F(\lambda_e) =S(\lambda_e)/4 \pi R^2$
via
\begin{equation}
f_\lambda(\lambda) = 
F\left(\frac{\lambda}{1+z}\right)\,T(\lambda,z) \frac{R^2}{(1+z)D_L^2(z)}\,,
 \label{eq:fl}
\end{equation}
where $S(\lambda_e)$ denotes the emitted spectral flux (as shown in Figure \ref{fig:spectra}), $T(\lambda,z)$ is the line of sight transmissivity of the interstellar medium and $D_L(z)$ is the luminosity distance to the source. The latter is given by
\begin{equation}
D_L(z) = (1+z) \frac{c}{H_0} \int_0^z \frac{dz'}{E(z')}
    \label{eq:Dl}\,,
    \vspace{0.1in}
\end{equation}
where $c$ is the speed of light, $H_0$  the Hubble constant, and 
\begin{equation}
E(z) = \sqrt{\Omega_m(1+z)^3 + \Omega_r(1+z)^4 + \Omega_\Lambda} 
\end{equation}
is the dimensionless Hubble parameter, with $ \Omega_m, \Omega_r, \Omega_\Lambda $ representing the matter, radiation, and dark energy density contributions, respectively. We assume a flat universe ($\Omega_\Lambda = 1 - \Omega_r - \Omega_m$) and compute $\Omega_r$ as the sum of the contributions from photons and neutrinos:
\begin{equation}
\Omega_r = \left (1 + \frac{7}{8} \left (\frac{4}{11} \right )^{4/3} N_\mathrm{eff} \right ) \frac{4\sigma}{c^3} \frac{T^4_\mathrm{CMB}}{\rho_c}
\end{equation}
where $T_\mathrm{CMB}$ is the temperature of the cosmic microwave background, 
$\sigma$ is the Stefan-Boltzmann constant,
$c$ the speed of light,
$N_\mathrm{eff}$ is the effective number of neutrino species and $\rho_c = 3H_0^2/8\pi G$ is the critical density. We adopt $H_0=68.22 \; \mathrm{km/s/Mpc}$ and $\Omega_m=0.3032$ \citep{ACT2025}, as well as $T_{\mathrm{CMB}}=2.725 \; \mathrm{K}$ and $N_\mathrm{eff}=3.04$.

Similarly to \citet{Larkin2023}, we model the line of sight transmissivity using the approximation from the numerical simulations by \citet{Meiksin2006},
\begin{equation}
    T(\lambda, z; z > 7) \approx 
\begin{cases}
0, & \text{if } \lambda \leq (1 + z)\,1215.67\,{\rm \AA} \\
1, & \text{otherwise},
\end{cases}
\end{equation}
and extend its validity also at redshifts below $z= 7$ given the blue cutoffs in the most optimal JWST bands for detections of Pop~III stars.
\begin{table}
\begin{tabular}{|c |c c|c c | } 
 \hline
 $z$ & Band 1 & $m_\mathrm{lim}$ & Band 2 & $m_\mathrm{lim}$ \\
 \hline
 3 & F070W & 29.68 & F090W & 29.83 \\ 
 7 & F115W & 29.97 & F150W & 30.22 \\ 
 12 & F200W & 30.39 & F322W2 & 30.70 \\ 
 17 & F277W & 30.43 & F356W & 30.47 \\ 
 \hline
\end{tabular}
\caption{Limiting magnitudes for the optimal pairs of NIRCAM bands at each redshift.}
\label{tab:limitingmags}
\end{table}

We determined the limiting magnitudes for all NIRCAM and MIRI bands using the JWST Exposure Time Calculator, specifically through the Pandeia package for Python (\citealt{Pontoppidan2016}). Limiting magnitudes were defined to be the greatest magnitude for which a flat spectrum would yield a signal-to-noise ratio above 3 under a 10 hour exposure, assuming a low background level measured at the ecliptic. The limiting magnitudes for relevant bands are provided in Table \ref{tab:limitingmags}.

To determine the optimal band for the purpose of detecting a Population III star, we computed the model's apparent magnitude in each band and used the one with the smallest difference from the band's  limiting magnitude. Likewise, we determined the optimal pair of bands for the purpose of measuring a star's color by computing the star's apparent magnitude in all non-overlapping pairs of bands, and using the pair with the smallest average difference from the limiting magnitudes of the bands.

Consistent with the findings of \citet{Larkin2023} (see their work for a detailed discussion), we observe that, at a given redshift, the majority of our stellar models share the same optimal band or pair of bands, largely independent of their initial mass. This result generally holds regardless of rotational velocity or evolutionary stage along the main sequence. The main exceptions are very massive stars approaching the end of their main sequence lifetimes, which exhibit a redward shift in their spectra due to the development of large convective cores. This redward shift generally becomes less pronounced as rotational velocity increases, but it can alter the optimal filter pair for massive stars with lower initial velocities, or models too massive for CHE to occur at any rotational velocity. However, given that such cases are rare, we adopt here a uniform set of bands for all models at each redshift, and then show in Appendix \ref{appendix:tams} some exceptions where the optimal band is different.

The lower-wavelength band in the optimal pair of bands was also generally the optimal band for detection at that redshift. While there are a few exceptions in which the higher-wavelength band is better for detecting very massive stars at IAMS and TAMS, these improvements are typically minor. Thus, the detectability of Pop III stars can be analyzed by focusing on their magnitudes in the lower-wavelength band.

Color-magnitude diagrams for all of our models at four different redshifts are provided in Figure \ref{fig:colormagnitude}. The observability limits under varying magnifications (shown with gray lines) are based only on the best band for detection, which is a conservative approximation that does not vary with color index. Our photometric results are consistent with those of \citet{Larkin2023} for non-rotating stars at ZAMS. However, there are some differences with rotation, which become especially marked as the star evolves.

For $\Omega=0.0$ and $\Omega=0.2$, our models typically evolve to cooler temperatures and progress redward over time in the color-magnitude diagram. On the other hand, the changes in color over time for rapidly rotating models depend on their mass. In lower mass models ($\lesssim 100 \; \mathrm{M}_\odot$), mixing remains efficient throughout the evolution, resulting in a steady increase of the temperature. At higher masses, the effect of increased mass loss and the associated angular momentum loss makes chemically homogeneous evolution harder to maintain, and we may no longer see the same monotonic increase in temperature at a given rotational velocity. Consequently, the temperature of high mass models at IAMS or TAMS could be lower than at ZAMS.

In addition to temperature, the color index is dependent on the composition of the atmosphere. In CHE and BSG models, rotational mixing and nuclear burning gradually replace surface hydrogen with helium, resulting in a reduction of hydrogen bound-free interactions. This usually shifts the star's color blueward. 

We find that rapidly-rotating stars with $M_\mathrm{init} \lesssim 350 \;\mathrm{M}_\odot$ do not display a significant color evolution between ZAMS and IAMS. Moreover, rapidly-rotating stars with $M_\mathrm{init} \lesssim 150 \;\mathrm{M}_\odot$ become bluer at TAMS. Stars too massive to achieve CHE at any rotational velocity generally evolve redward, even at high rotational velocities.

Ultimately, the color index of a star with a given mass is noticeably affected by its rotational speed and age; however, with the exception of massive models near the end of main sequence, all stars at a given redshift are still best observed using the same NIRCAM band. Rotation and age are also found to affect the star's apparent magnitude, especially in the case of more massive models. Some of our most massive models were found to be detectable at magnifications below $\mu = 10^2$ during IAMS, even for redshifts up to $z=12$. If Pop~III stars were to exist  down to redshifts of $z\sim 3$, the most massive ones $\sim 800 \;\mathrm{M}_\odot$ could almost be directly detected without lensing in the MIRI bands of JWST (cfr. Fig.~\ref{fig:colormagnitude_hi} in Appendix \ref{appendix:tams}).

\section{Pop~III in multiple star systems}\label{sec:clusters}
The number of Pop~III stars forming within a single cluster remains an open question. While early studies suggested that massive Pop~III stars predominantly formed in isolation (e.g., \citealt{Abel2002, Bromm2002}), more recent simulations have shown that angular momentum in the collapsing gas cloud can lead to the formation of a self-gravitating protostellar disc. Such discs are prone to fragmentation, potentially resulting in the formation of multiple stars (e.g., \citealt{Clark2008, Turk2009,Johnson2010,greif2011,stacy2012, stacy2016,Visbal2017,wollenberg2020}).

The exact number of protostars remains uncertain due to the complex and chaotic nature of early star formation processes. However, 
the numerical studies generally suggest that Pop III star clusters  can consist of a few to several dozen stars, depending on factors such as gas dynamics, turbulence, and radiative feedback.  
The presence of larger clusters is also constrained by the lack of observations of Pop~III stars in the Milky Way \citep{ishiyama2016}, albeit this constraint is degenerate with respect to the minimum formation mass of Pop~III stars, which is rather uncertain \citep{hartwig2015}.
There are also several uncertainties regarding the correct form of the initial mass function (IMF) for Population III stars (\citealt{GesseyJones2022}), though a common approach motivated by hydrodynamic models is to adopt a top-heavy IMF with a maximum mass of $150 \; \mathrm{M}_\odot$ (e.g. \citealt{Wang2022}, \citealt{Wu2025}).

Given these uncertainties, we proceed by examining the impact of multiple gravitationally bound stars by treating the total number of stars as a free parameter. The masses of Population III stars are drawn from the initial mass function (IMF) derived from the recent simulations by \citet{reinoso2025}, conducted using the ENZO code \citep{Bryan2014}. They found the IMF, $\xi(M)$, to be  approximated by:
\begin{equation}
\xi(M) \propto M^{-\alpha_i},
\label{eq:IMF}
\end{equation}
\noindent
where
\[
\begin{aligned}
\alpha_1 &= \phantom{-}0.66, & \quad 0.4 \leq \frac{M}{M_\odot} < 3.0, \\
\alpha_2 &= -0.15, & \quad 3.0 \leq \frac{M}{M_\odot} < 10.0, \\
\alpha_3 &= \phantom{-}0.83, & \quad 10.0 \leq \frac{M}{M_\odot} < 20.0, \\
\alpha_4 &= \phantom{-}1.75, & \quad 20.0 \leq \frac{M}{M_\odot} < 150.0.
\end{aligned}
\]

In the particular example that we study here, since we adopt the IMF from \citet{reinoso2025}, we model stars only within the mass range defined by their simulation at the point of cluster formation, corresponding to the termination of the ENZO run. It is important to note, however, that the high-mass end of the IMF is expected to grow further through continued accretion and stellar collisions \citep{reinoso2025}.

Given the short lifetime of evolved stars compared to the main sequence, we restrict our focus to clusters for which all the stars are on the main sequence. In particular, we assume all of the stars are simultaneously at ZAMS, which will provide a lower bound for the cluster's luminosity. We also assume all of the stars are nonrotating; aside from nonrotating stars being computationally faster to evolve, our results from the above sections indicate the spectra of nonrotating stars at ZAMS are generally similar to those of rotating stars, as there is not enough time for significant mass loss or dredging of metals from the core.

We consider clusters of several multiplicities ranging from $n=5$ to 80.\footnote{The simulation by \citet{reinoso2025} identified 68 particles at the end.} For each multiplicity, we randomly sample star masses using the IMF, compute the photometry of the individual stars, and then sum their fluxes to find the total photometry of the cluster. We repeat this for a total of 10,000 trials per cluster. Summary statistics across these trials for each cluster are provided in Table~\ref{tab:cluster} in Appendix \ref{appendix:tables}.

\begin{figure*}[ht!]
\centering
\includegraphics[width=\textwidth]{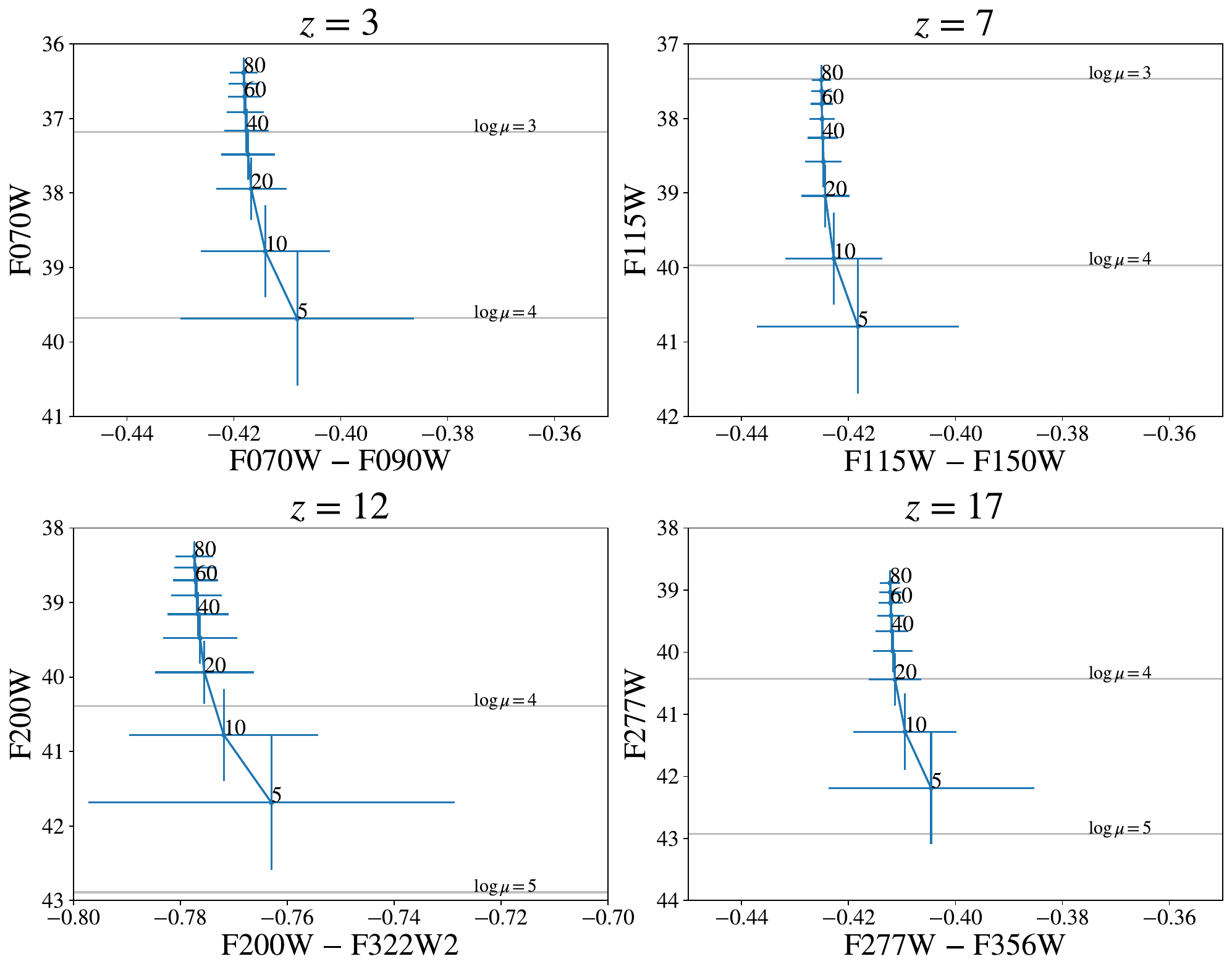}
\caption{Color magnitude diagrams for clusters of varying multiplicities. 
The plotted values for each cluster were obtained by summing the fluxes for the cluster's individual stars, computing the total magnitude and color index for the cluster, and then averaging the cluster's magnitude and color across 10,000 trials. 
The error bars indicate the $1\sigma$ dispersion.
In contrast to Figure \ref{fig:colormagnitude}, the numbers beside the points denote the cluster's multiplicity rather than its mass.}
\label{fig:cluster_CMD}
\end{figure*}

Color-magnitude diagrams for our clusters are shown in Figure \ref{fig:cluster_CMD}, using the same pairs of bands derived for individual stars (see Table ~\ref{tab:limitingmags}). Note that the relationship between a cluster's average mass and photometry across 10,000 trials is not the same as the mass-photometry relationship seen for individual stars in Figure \ref{fig:colormagnitude}. Our clusters were found to be consistently dimmer than individual stars of the same total mass and evolutionary phase, with the difference becoming increasingly significant for larger clusters. Even so, the largest of our clusters were found to be visible at magnifications below $\mu \sim 10^3$ at redshifts up to $z=7$. Thus, they may offer another possibility for detecting Population III stars under strong lensing, without relying on the formation of large individual stars. Additionally, while our computations were restricted to ZAMS, it may be the case that more evolved clusters would be visible at lower magnifications, as was the case for individual stars.

\section{Connection to high-$Z$ GRBs}\label{sec:GRB}

As discussed above, Pop~III stars can form in excess of $100 M_{\odot}$ \citep{Abel2002,Bromm2002}, and their low metallicity can suppress mass loss throughout the star's lifetime, allowing these early stars to maintain angular momentum up until stellar collapse (see e.g. \citealt{stacy2013}). This feature makes Pop~III stars a promising candidate for the production of GRBs.
In fact, GRBs from Pop~III stars have been suggested as a probe of the high-$z$ universe \citep{Lamb2000,Ciardi2000,Gou2004,Ioka2005,Mesinger2005,Inoue2007,Toma2011,desouza2013}, and are thus a subject of continued interest. Observationally, they may appear to belong to a class of particularly long-lived GRBs known as ultra-long gamma-ray bursts (ULGRBs) that can last several thousand seconds \citep{levan_swift_2015}. While other, special evolutionary stellar paths have been proposed to produce ULGRBs (e.g. \citealt{perna2018}), the massive stellar envelops present in Pop~III stars can provide a source for powerful accretion that can last much longer than stars in the standard collapsar model of LGRBs, while the angular momentum in the rapidly spinning black hole remnants could efficiently launch ultra-relativistic jets capable of producing ULGRBs.  Any jet produced in this manner would, however, face the obstacle of successfully breaking out of the remaining stellar envelope that would likely extend to very large radii. Table~\ref{tab:table1}, for example, quotes TAMS radii in the thousands of solar radii for the most massive Pop~III stars that we analyzed. Extended progenitor stars could also affect the emission properties of the bursts, for example by causing a larger mass entrainment that would push the photosphere to larger radii, softening and thermalizing the prompt emission spectrum. 

It has been shown, analytically, that jets launched from collapsing Pop III stars can break out from their stellar envelope \citep{suwa_can_2010}.  Additional work has reached similar conclusions resulting from numerical simulations of relativistic jets launched self-consistently through accretion of the stellar envelope \citep{nagakura_population_2012}.  
The models generated for this paper constitute an additional step at providing a better understanding of the connection between Pop III stars and GRBs. A detailed calculation of the most realistic pre-explosion stellar models is indeed required to pin down the likelihood of jet formation and propagation in these stars. Such a detailed model will also allow for a robust prediction of the observational characteristics of very high-redshift events, possibly leading to their identification in transient surveys. To fully exploit this potential, we plan to follow up with an end-to-end hydrodynamical and radiation transfer study of the jet launched by the central engine in a forthcoming work (Walker et al., in preparation).

\section{Summary}\label{sec:summary}

Motivated by the prospect of direct observation with JWST of the first stars in the Universe, the $\sim$ zero-metallicity so-called Pop~III stars,  here we have performed new evolutionary calculations to characterize their expected observational signatures. In particular, building on previous work on non-rotating, ZAMS stars  \citep{Larkin2023}, we have studied the effect of rotation on the expected spectra, given that these stars are expected to be generally fast rotators \citep{stacy2013}. We further evolved these stars past ZAMS, to study the spectral changes as the stars age. Although this phase is brief, the spectral changes can be significant.
Finally, we considered the observability and observational signatures of a cluster of Pop~III stars. 

Our main results can be summarized as follows:
\begin{itemize}
    \item The optimal bands for detecting Pop~III stars with JWST, as well as the optimal pairs of bands for measuring their colors, are determined by    
    a combination of their intrinsic spectrum, their redshift, and the JWST
    sensitivity function in the various bands. We generally find that the optimal filters are the same for all stars at the same redshift, and vary between 
    F070W at $z=3$ to F277W at $z=17$.
    However, there are exceptions for massive ($\gtrsim 100 \; \mathrm{M}_\odot$), slowly-rotating stars near the end of the main sequence, which progress far redward, and for which redder JWST filters are the optimal ones.
    \item Pop III stars following the RSG track become significantly brighter over time, with their apparent magnitudes in the optimal bands typically decreasing by 1-2 over the course of main-sequence evolution. More rapidly rotating models on the BSG and CHE tracks tend to have less pronounced changes in magnitude over time.
    \item We find that accounting for rotation and evolution lowers the magnification threshold needed to detect Pop III stars. This is particularly substantial for very massive stars with $M\gtrsim 500 \;\mathrm{M}_\odot$, which can become very bright post-ZAMS phase. Such stars could be detected with moderate lensing amplifications of $\mu \lesssim 10^2$ at IAMS and 
     $\mu \lesssim 10$ at TAMS,
     for distances $z\lesssim 7$. Less massive and more distant stars will still require very favorable lensing ($\mu \gtrsim 10^3$) to be detected.
    \item Our simulated Pop~III star clusters have similar trial-averaged color indices at a given redshift, regardless of cluster size. Moreover, the optimal bands for observing clusters are the same as the optimal bands for observing individual stars at the same redshift.
    \item Clusters of sizes $n=$ 5, 10, 20, and 30 are found to be significantly dimmer compared to single stars of equivalent total mass. Large clusters of size $n\sim70$ are found to be comparable in magnitude to the largest stable Pop~III mass ($\sim 820 \; \mathrm{M}_\odot$) predicted by \citep{Larkin2023}, and can potentially be detected at magnifications below $\mu \sim 10^3$.
    \item The effects of rotation and evolution were not considered for clusters. However, the assumption of all stars in a cluster being at ZAMS is expected to serve as a lower bound for the magnitude, and it is possible that evolved clusters may further lower the magnification threshold for detecting Pop~III stars.
\end{itemize}

Looking ahead, continued refinement of Pop III spectral models — including the inclusion of binary evolution, mass loss, and nebular emission — will be essential for improving identification strategies. As JWST deep field campaigns and time-domain lensing surveys expand, the likelihood of detecting highly magnified Pop III stars or compact clusters will grow. 

Synergistic observations with gravitational lensing reconstructions, high-resolution spectroscopy, and transient monitoring may enable the first robust constraints on Pop~III stellar properties. Ultimately, the combination of advanced theoretical modeling and next-generation observational campaigns promises to transform Pop~III science from speculative theory to empirical exploration during the JWST era and beyond.

\section*{Acknowledgements}

R.P. and J.H. gratefully acknowledge support by NSF award AST-2006839 and NASA award 80NSSC25K7554.
The Center for Computational Astrophysics at the Flatiron Institute is supported by the Simons Foundation

\software{\texttt{MESA} \citep{paxton2011,paxton2013,paxton2015,paxton2018,paxton2019},
\texttt{Astropy} \citep{Astropy2013}
\texttt{Matplotlib} \citep{Hunter2007},
\texttt{BasicATLAS} \citep{Gerasimov2022},
\texttt{Pandeia} \citep{Pontoppidan2016}.}

\appendix

\section{Model outputs}\label{appendix:tables}

\noindent
We report in Table~\ref{tab:table1} and Table~\ref{tab:table2} the most relevant \texttt{MESA} inputs and outputs for the evolutionary models that we computed. The statistics for the Pop~III star clusters are reported in Table~\ref{tab:cluster}.

\begin{longtable}{c c c | c c c c c c c c c c}
 \caption{\texttt{MESA} outputs for the models used, where ``init" denotes the initial parameters. $M$: mass; $v_{\mathrm{init}} / v_K$: ratio of initial rotational velocity with Keplerian velocity; $R$: radius of photosphere; $L$: luminosity. The Mode denotes the evolutionary track the model follows, as defined in Section \ref{sec:methods}} \\
 \hline
 $M_{\mathrm{init}}$ & $\Omega_{\mathrm{init}}$ & $v_{\mathrm{init}} / v_K$ & $M_{\mathrm{ZAMS}}$ & $R_{\mathrm{ZAMS}}$ & $L_\mathrm{ZAMS}$ & $M_{\mathrm{IAMS}}$ & $R_{\mathrm{IAMS}}$ & $L_\mathrm{IAMS}$ & $M_{\mathrm{TAMS}}$ & $R_{\mathrm{TAMS}}$ & $L_\mathrm{TAMS}$ & Mode \\ ($M_\odot$) & & & ($M_\odot$) & ($R_\odot$) & ($L_\odot$) & ($M_\odot$) & ($R_\odot$) & ($L_\odot$) & ($M_\odot$) & ($R_\odot$) & ($L_\odot$) & \\ \hline\endfirsthead
 \caption{(Continued)}\\
 \hline
 $M_{\mathrm{init}}$ & $\Omega_{\mathrm{init}}$ & $v_{\mathrm{init}} / v_K$ & $M_{\mathrm{ZAMS}}$ & $R_{\mathrm{ZAMS}}$ & $L_\mathrm{ZAMS}$ & $M_{\mathrm{IAMS}}$ & $R_{\mathrm{IAMS}}$ & $L_\mathrm{IAMS}$ & $M_{\mathrm{TAMS}}$ & $R_{\mathrm{TAMS}}$ & $L_\mathrm{TAMS}$ & Mode \\ ($M_\odot$) & & & ($M_\odot$) & ($R_\odot$) & ($L_\odot$) & ($M_\odot$) & ($R_\odot$) & ($L_\odot$) & ($M_\odot$) & ($R_\odot$) & ($L_\odot$) & \\
 \hline
 \endhead
 \label{tab:table1}
10 & 0.00 & 0.00 & 10.0 & 1.39 & 7080 & 10.0 & 1.94 & 13247 & 10.0 & 3.22 & 20273 & RSG\\
10 & 0.20 & 0.19 & 10.0 & 1.42 & 6872 & 10.0 & 2.04 & 13555 & 10.0 & 3.61 & 21368 & RSG\\
10 & 0.40 & 0.38 & 10.0 & 1.61 & 6586 & 10.0 & 1.60 & 33815 & 9.9 & 1.39 & 121588 & CHE\\
10 & 0.60 & 0.57 & 9.9 & 1.75 & 6026 & 9.8 & 1.70 & 31030 & 9.7 & 1.37 & 114362 & CHE\\
\hline
25 & 0.00 & 0.00 & 25.0 & 1.89 & 89070 & 25.0 & 3.84 & 151346 & 25.0 & 8.29 & 219075 & RSG\\
25 & 0.20 & 0.17 & 25.0 & 1.92 & 83460 & 25.0 & 4.03 & 152022 & 25.0 & 10.62 & 226307 & RSG\\
25 & 0.40 & 0.35 & 25.0 & 2.37 & 78278 & 25.0 & 2.78 & 264766 & 24.6 & 2.15 & 615348 & CHE\\
25 & 0.60 & 0.53 & 24.4 & 2.27 & 73987 & 24.4 & 2.86 & 245157 & 23.9 & 2.18 & 575220 & CHE\\
\hline
50 & 0.00 & 0.00 & 50.0 & 2.83 & 403226 & 50.0 & 6.00 & 616549 & 50.0 & 19.39 & 823951 & RSG\\
50 & 0.20 & 0.15 & 50.0 & 2.88 & 394955 & 50.0 & 6.14 & 682883 & 49.6 & 30.77 & 946679 & RSG\\
50 & 0.40 & 0.31 & 50.0 & 3.63 & 381211 & 50.0 & 4.12 & 896381 & 48.7 & 3.25 & 1675890 & CHE\\
50 & 0.60 & 0.48 & 48.9 & 3.39 & 354722 & 48.8 & 4.10 & 859867 & 46.8 & 3.25 & 1586253 & CHE\\
\hline
75 & 0.00 & 0.00 & 75.0 & 3.51 & 837677 & 75.0 & 7.95 & 1233067 & 75.0 & 39.19 & 1565238 & RSG\\
75 & 0.20 & 0.14 & 75.0 & 3.59 & 832028 & 75.0 & 7.57 & 1341707 & 74.2 & 38.66 & 1769882 & RSG\\
75 & 0.40 & 0.29 & 75.0 & 4.15 & 811021 & 75.0 & 5.28 & 1710063 & 72.6 & 4.06 & 2866040 & CHE\\
75 & 0.60 & 0.44 & 73.5 & 4.24 & 755390 & 73.0 & 5.08 & 1615721 & 69.9 & 3.96 & 2734434 & CHE\\
\hline
100 & 0.00 & 0.00 & 100.0 & 4.12 & 1363587 & 100.0 & 9.72 & 1924702 & 100.0 & 85.72 & 2377845 & RSG\\
100 & 0.20 & 0.14 & 100.0 & 4.20 & 1345041 & 100.0 & 9.88 & 2052869 & 98.7 & 81.21 & 2567404 & RSG\\
100 & 0.40 & 0.27 & 100.0 & 4.69 & 1305948 & 100.0 & 6.41 & 2570211 & 96.0 & 4.91 & 4080814 & CHE\\
100 & 0.60 & 0.42 & 98.1 & 4.92 & 1237407 & 97.2 & 5.92 & 2453660 & 92.5 & 4.73 & 3917175 & CHE\\
\hline
150 & 0.00 & 0.00 & 150.0 & 5.14 & 2550344 & 150.0 & 13.00 & 3422201 & 150.0 & 3499.91 & 3914822 & RSG\\
150 & 0.20 & 0.12 & 150.0 & 5.24 & 2533775 & 150.0 & 12.46 & 3665725 & 147.5 & 1144.36 & 4408368 & RSG\\
150 & 0.40 & 0.25 & 150.0 & 5.75 & 2477514 & 150.0 & 8.02 & 4465058 & 139.1 & 8.93 & 6207745 & CHE\\
150 & 0.60 & 0.39 & 147.8 & 6.10 & 2344246 & 145.2 & 7.42 & 4272776 & 135.3 & 7.25 & 6025887 & CHE\\
\hline
200 & 0.00 & 0.00 & 200.0 & 6.06 & 3896440 & 200.0 & 16.68 & 5038506 & 200.0 & 4441.17 & 5919245 & RSG\\
200 & 0.20 & 0.12 & 200.0 & 6.13 & 3847826 & 200.0 & 19.27 & 5349697 & 195.7 & 4439.89 & 6301556 & RSG\\
200 & 0.40 & 0.24 & 200.0 & 6.65 & 3757720 & 191.6 & 14.01 & 5868902 & 184.9 & 872.54 & 7204064 & RSG\\
200 & 0.60 & 0.36 & 197.5 & 7.06 & 3581955 & 193.5 & 8.63 & 6217570 & 177.3 & 10.82 & 8185642 & CHE\\
\hline
350 & 0.00 & 0.00 & 350.0 & 8.23 & 8226644 & 350.0 & 29.78 & 10163495 & 350.0 & 6051.26 & 11500158 & RSG\\
350 & 0.20 & 0.10 & 350.0 & 8.35 & 8180322 & 347.2 & 35.60 & 10281895 & 342.6 & 5972.24 & 11672336 & RSG\\
350 & 0.40 & 0.21 & 350.0 & 8.92 & 8047551 & 331.2 & 30.29 & 10695214 & 327.3 & 5822.72 & 12314343 & RSG\\
350 & 0.60 & 0.32 & 346.9 & 9.60 & 7706547 & 336.8 & 11.80 & 12370917 & 304.8 & 114.05 & 15187560 & BSG\\
\hline
450 & 0.00 & 0.00 & 450.0 & 9.49 & 11315800 & 450.0 & 39.55 & 13672411 & 450.0 & 6863.36 & 15210256 & RSG\\
450 & 0.20 & 0.10 & 450.0 & 9.64 & 11255237 & 444.6 & 50.41 & 13769113 & 440.5 & 6815.83 & 15456728 & RSG\\
450 & 0.40 & 0.20 & 450.0 & 10.22 & 11085058 & 424.1 & 51.76 & 14518678 & 421.2 & 6668.10 & 16468279 & RSG\\
450 & 0.60 & 0.30 & 446.9 & 10.90 & 10649884 & 412.8 & 28.41 & 14167828 & 404.4 & 6430.37 & 16027681 & RSG\\
\hline
500 & 0.00 & 0.00 & 500.0 & 10.08 & 12896891 & 500.0 & 47.33 & 15471198 & 500.0 & 7246.72 & 17177456 & RSG\\
500 & 0.20 & 0.09 & 500.0 & 10.20 & 12835340 & 492.3 & 62.51 & 15571893 & 489.4 & 7227.17 & 17435631 & RSG\\
500 & 0.40 & 0.19 & 500.0 & 10.83 & 12642107 & 473.9 & 50.39 & 15725850 & 470.4 & 7063.82 & 17515493 & RSG\\
500 & 0.60 & 0.29 & 496.8 & 11.55 & 12166092 & 462.3 & 26.77 & 15776853 & 450.0 & 6832.69 & 17707525 & RSG\\
\hline
550 & 0.00 & 0.00 & 550.0 & 10.61 & 14499167 & 550.0 & 53.59 & 17250499 & 550.0 & 7570.12 & 19053384 & RSG\\
550 & 0.20 & 0.09 & 550.0 & 10.76 & 14429378 & 541.3 & 66.18 & 17287628 & 538.2 & 7520.72 & 19231913 & RSG\\
550 & 0.40 & 0.19 & 550.0 & 11.34 & 14233511 & 518.4 & 76.17 & 17596142 & 516.2 & 7419.69 & 19622712 & RSG\\
550 & 0.60 & 0.29 & 546.7 & 12.17 & 13701844 & 501.9 & 38.26 & 17438113 & 495.3 & 7225.68 & 19511913 & RSG\\
\hline
600 & 0.00 & 0.00 & 600.0 & 11.11 & 16114922 & 600.0 & 64.35 & 19067230 & 600.0 & 7856.33 & 20765230 & RSG\\
600 & 0.20 & 0.09 & 600.0 & 11.24 & 16042071 & 589.4 & 99.64 & 19260573 & 587.3 & 7892.93 & 21531784 & RSG\\
600 & 0.40 & 0.18 & 600.0 & 11.93 & 15822186 & 564.5 & 101.10 & 19365693 & 563.2 & 7711.96 & 21514807 & RSG\\
600 & 0.60 & 0.28 & 596.6 & 12.81 & 15276281 & 547.6 & 47.01 & 19106668 & 541.3 & 7571.98 & 21230896 & RSG\\
\hline
650 & 0.00 & 0.00 & 650.0 & 11.61 & 17744221 & 650.0 & 73.77 & 20863038 & 650.0 & 8212.35 & 22869493 & RSG\\
650 & 0.20 & 0.09 & 650.0 & 11.75 & 17666706 & 639.3 & 86.91 & 20834832 & 636.5 & 8076.58 & 22982978 & RSG\\
650 & 0.40 & 0.18 & 650.0 & 12.39 & 17438109 & 612.5 & 92.23 & 21187741 & 610.4 & 7979.26 & 23446418 & RSG\\
650 & 0.60 & 0.28 & 646.5 & 13.38 & 16843966 & 589.7 & 63.02 & 20835249 & 586.3 & 7839.90 & 23131750 & RSG\\
\hline
700 & 0.00 & 0.00 & 700.0 & 12.12 & 19386884 & 700.0 & 85.79 & 22682858 & 700.0 & 8496.53 & 24792637 & RSG\\
700 & 0.20 & 0.09 & 700.0 & 12.25 & 19304296 & 686.1 & 194.28 & 22787874 & 684.9 & 8434.10 & 25082972 & RSG\\
700 & 0.40 & 0.17 & 700.0 & 12.90 & 19059083 & 659.1 & 185.47 & 22845127 & 658.1 & 8289.72 & 25236157 & RSG\\
700 & 0.60 & 0.27 & 696.4 & 13.94 & 18431968 & 637.8 & 57.33 & 22449629 & 632.2 & 8069.50 & 24739219 & RSG\\
\hline
750 & 0.00 & 0.00 & 750.0 & 12.61 & 21039752 & 750.0 & 103.34 & 24503318 & 750.0 & 8770.02 & 26741386 & RSG\\
750 & 0.20 & 0.08 & 750.0 & 12.72 & 20952359 & 735.2 & 199.55 & 24535424 & 733.7 & 8663.37 & 27051692 & RSG\\
750 & 0.40 & 0.17 & 750.0 & 13.39 & 20692864 & 707.5 & 120.94 & 24628739 & 705.6 & 8554.95 & 27194814 & RSG\\
750 & 0.60 & 0.27 & 746.4 & 14.51 & 20039428 & 685.3 & 59.22 & 24131062 & 677.9 & 8323.31 & 26506145 & RSG\\
\hline
800 & 0.00 & 0.00 & 800.0 & 13.12 & 22700586 & 800.0 & 120.56 & 26338077 & 800.0 & 9031.92 & 28663859 & RSG\\
800 & 0.20 & 0.08 & 800.0 & 13.19 & 22610032 & 783.9 & 392.43 & 26429702 & 782.6 & 8891.02 & 28937028 & RSG\\
800 & 0.40 & 0.17 & 800.0 & 13.87 & 22335200 & 753.3 & 400.52 & 26660309 & 752.1 & 8818.46 & 29373293 & RSG\\
800 & 0.60 & 0.26 & 796.3 & 14.99 & 21637562 & 724.7 & 130.91 & 26235259 & 722.6 & 8623.64 & 28934237 & RSG\\
\hline
\end{longtable}

\begin{longtable}{c c | c c c c c c c c c c}
 \caption{Additional \texttt{MESA} outputs. $Y_\mathrm{s}$: surface helium mass fraction; $Z_\mathrm{s}$: surface metal mass fraction; $T_\mathrm{eff}$: effective temperature.} \\
 \hline
$M_{\mathrm{init}}$ & $\Omega_{\mathrm{init}}$ & $T_\mathrm{eff,ZAMS}$ & $Y_\mathrm{s,ZAMS}$ & $Z_\mathrm{s,ZAMS}$ & $T_\mathrm{eff,IAMS}$ & $Y_\mathrm{s,IAMS}$ & $Z_\mathrm{s,IAMS}$ & $T_\mathrm{eff,TAMS}$ & $Y_\mathrm{s,TAMS}$ & $Z_\mathrm{s,TAMS}$ \\ ($M_\odot$) & & (K) & & $(10^{-8})$ & (K) & & $(10^{-8})$ & (K) & & $(10^{-8})$ \\ \hline\endfirsthead
 \caption{(Continued)}\\
 \hline
$M_{\mathrm{init}}$ & $\Omega_{\mathrm{init}}$ & $T_\mathrm{eff,ZAMS}$ & $Y_\mathrm{s,ZAMS}$ & $Z_\mathrm{s,ZAMS}$ & $T_\mathrm{eff,IAMS}$ & $Y_\mathrm{s,IAMS}$ & $Z_\mathrm{s,IAMS}$ & $T_\mathrm{eff,TAMS}$ & $Y_\mathrm{s,TAMS}$ & $Z_\mathrm{s,TAMS}$ \\ ($M_\odot$) & & (K) & & $(10^{-8})$ & (K) & & $(10^{-8})$ & (K) & & $(10^{-8})$ \\
 \hline
 \endhead
 \label{tab:table2}
10 & 0.00 & 44866 & 0.24 & 0.00 & 44424 & 0.24 & 0.00 & 38373 & 0.24 & 0.00\\
10 & 0.20 & 44098 & 0.24 & 0.00 & 43590 & 0.25 & 0.00 & 36714 & 0.25 & 0.00\\
10 & 0.40 & 41001 & 0.24 & 0.00 & 61820 & 0.66 & 0.09 & 91472 & 0.97 & 2.38\\
10 & 0.60 & 38496 & 0.24 & 0.00 & 58840 & 0.64 & 0.08 & 90612 & 0.97 & 2.21\\
\hline
25 & 0.00 & 72445 & 0.24 & 0.00 & 58063 & 0.24 & 0.00 & 43376 & 0.24 & 0.00\\
25 & 0.20 & 70880 & 0.24 & 0.00 & 56765 & 0.25 & 0.01 & 38628 & 0.25 & 0.01\\
25 & 0.40 & 62778 & 0.24 & 0.00 & 78591 & 0.67 & 0.29 & 110332 & 0.99 & 8.66\\
25 & 0.60 & 63144 & 0.24 & 0.00 & 75890 & 0.66 & 0.28 & 107734 & 0.99 & 7.44\\
\hline
50 & 0.00 & 86537 & 0.24 & 0.00 & 66042 & 0.24 & 0.00 & 39494 & 0.24 & 0.00\\
50 & 0.20 & 85203 & 0.24 & 0.00 & 66982 & 0.31 & 0.07 & 32457 & 0.31 & 0.08\\
50 & 0.40 & 75244 & 0.24 & 0.00 & 87496 & 0.66 & 0.48 & 115283 & 0.99 & 8.34\\
50 & 0.60 & 76470 & 0.24 & 0.00 & 86777 & 0.67 & 0.49 & 113713 & 0.97 & 5.78\\
\hline
75 & 0.00 & 93160 & 0.24 & 0.00 & 68236 & 0.24 & 0.00 & 32614 & 0.24 & 0.00\\
75 & 0.20 & 91952 & 0.24 & 0.00 & 71385 & 0.33 & 0.12 & 33857 & 0.37 & 0.17\\
75 & 0.40 & 84983 & 0.24 & 0.00 & 90830 & 0.67 & 0.62 & 117931 & 0.98 & 8.41\\
75 & 0.60 & 82685 & 0.24 & 0.00 & 91340 & 0.67 & 0.62 & 117888 & 0.98 & 9.45\\
\hline
100 & 0.00 & 97217 & 0.24 & 0.00 & 68944 & 0.24 & 0.00 & 24481 & 0.24 & 0.00\\
100 & 0.20 & 95958 & 0.24 & 0.00 & 69519 & 0.31 & 0.13 & 25638 & 0.34 & 0.15\\
100 & 0.40 & 90079 & 0.24 & 0.00 & 91267 & 0.66 & 0.68 & 117021 & 0.97 & 5.97\\
100 & 0.60 & 86761 & 0.24 & 0.00 & 93892 & 0.67 & 0.72 & 118112 & 0.97 & 7.18\\
\hline
150 & 0.00 & 101736 & 0.24 & 0.00 & 68848 & 0.24 & 0.00 & 4340 & 0.24 & 0.00\\
150 & 0.20 & 100621 & 0.24 & 0.00 & 71540 & 0.34 & 0.19 & 7818 & 0.37 & 0.25\\
150 & 0.40 & 95534 & 0.24 & 0.00 & 93706 & 0.66 & 0.81 & 96440 & 0.86 & 2.18\\
150 & 0.60 & 91442 & 0.24 & 0.00 & 96355 & 0.67 & 0.84 & 106210 & 0.90 & 2.94\\
\hline
200 & 0.00 & 104131 & 0.24 & 0.00 & 66956 & 0.24 & 0.00 & 4272 & 0.24 & 0.00\\
200 & 0.20 & 103252 & 0.24 & 0.00 & 63231 & 0.31 & 0.17 & 4340 & 0.34 & 0.21\\
200 & 0.40 & 98551 & 0.24 & 0.00 & 75906 & 0.44 & 0.38 & 10123 & 0.48 & 0.46\\
200 & 0.60 & 94488 & 0.24 & 0.00 & 98121 & 0.68 & 0.95 & 93874 & 0.85 & 2.24\\
\hline
350 & 0.00 & 107750 & 0.24 & 0.00 & 59716 & 0.24 & 0.00 & 4321 & 0.24 & 0.00\\
350 & 0.20 & 106798 & 0.24 & 0.00 & 54776 & 0.26 & 0.08 & 4366 & 0.30 & 0.18\\
350 & 0.40 & 102944 & 0.24 & 0.00 & 59979 & 0.31 & 0.23 & 4481 & 0.47 & 0.51\\
350 & 0.60 & 98140 & 0.24 & 0.00 & 99657 & 0.68 & 1.11 & 33740 & 0.79 & 1.84\\
\hline
450 & 0.00 & 108643 & 0.24 & 0.00 & 55809 & 0.24 & 0.00 & 4351 & 0.24 & 0.00\\
450 & 0.20 & 107689 & 0.24 & 0.00 & 49523 & 0.25 & 0.06 & 4384 & 0.32 & 0.24\\
450 & 0.40 & 104159 & 0.24 & 0.00 & 49522 & 0.33 & 0.27 & 4503 & 0.41 & 0.43\\
450 & 0.60 & 99884 & 0.24 & 0.00 & 66438 & 0.41 & 0.42 & 4554 & 0.48 & 0.55\\
\hline
500 & 0.00 & 108945 & 0.24 & 0.00 & 52619 & 0.24 & 0.00 & 4365 & 0.24 & 0.00\\
500 & 0.20 & 108181 & 0.24 & 0.00 & 45859 & 0.26 & 0.08 & 4387 & 0.29 & 0.17\\
500 & 0.40 & 104561 & 0.24 & 0.00 & 51205 & 0.31 & 0.24 & 4443 & 0.37 & 0.34\\
500 & 0.60 & 100298 & 0.24 & 0.00 & 70310 & 0.41 & 0.44 & 4530 & 0.48 & 0.59\\
\hline
550 & 0.00 & 109352 & 0.24 & 0.00 & 50814 & 0.24 & 0.00 & 4383 & 0.24 & 0.00\\
550 & 0.20 & 108435 & 0.24 & 0.00 & 45752 & 0.25 & 0.08 & 4408 & 0.32 & 0.24\\
550 & 0.40 & 105279 & 0.24 & 0.00 & 42835 & 0.29 & 0.20 & 4460 & 0.39 & 0.40\\
550 & 0.60 & 100671 & 0.24 & 0.00 & 60303 & 0.38 & 0.39 & 4513 & 0.44 & 0.51\\
\hline
600 & 0.00 & 109729 & 0.24 & 0.00 & 47546 & 0.24 & 0.00 & 4396 & 0.24 & 0.00\\
600 & 0.20 & 108956 & 0.24 & 0.00 & 38308 & 0.27 & 0.12 & 4426 & 0.35 & 0.30\\
600 & 0.40 & 105385 & 0.24 & 0.00 & 38081 & 0.28 & 0.17 & 4476 & 0.40 & 0.44\\
600 & 0.60 & 100804 & 0.24 & 0.00 & 55656 & 0.37 & 0.38 & 4503 & 0.41 & 0.44\\
\hline
650 & 0.00 & 109922 & 0.24 & 0.00 & 45418 & 0.24 & 0.00 & 4405 & 0.24 & 0.00\\
650 & 0.20 & 109157 & 0.24 & 0.00 & 41831 & 0.26 & 0.09 & 4447 & 0.36 & 0.34\\
650 & 0.40 & 105954 & 0.24 & 0.00 & 40777 & 0.31 & 0.24 & 4496 & 0.43 & 0.50\\
650 & 0.60 & 101102 & 0.24 & 0.00 & 49124 & 0.35 & 0.34 & 4521 & 0.44 & 0.52\\
\hline
700 & 0.00 & 110001 & 0.24 & 0.00 & 43007 & 0.24 & 0.00 & 4419 & 0.24 & 0.00\\
700 & 0.20 & 109328 & 0.24 & 0.00 & 28612 & 0.25 & 0.06 & 4448 & 0.31 & 0.25\\
700 & 0.40 & 106182 & 0.24 & 0.00 & 29301 & 0.28 & 0.17 & 4493 & 0.40 & 0.43\\
700 & 0.60 & 101293 & 0.24 & 0.00 & 52475 & 0.36 & 0.36 & 4532 & 0.45 & 0.56\\
\hline
750 & 0.00 & 110091 & 0.24 & 0.00 & 39948 & 0.24 & 0.00 & 4432 & 0.24 & 0.00\\
750 & 0.20 & 109475 & 0.24 & 0.00 & 28757 & 0.25 & 0.07 & 4472 & 0.37 & 0.38\\
750 & 0.40 & 106374 & 0.24 & 0.00 & 36975 & 0.31 & 0.25 & 4506 & 0.41 & 0.48\\
750 & 0.60 & 101367 & 0.24 & 0.00 & 52569 & 0.35 & 0.34 & 4540 & 0.46 & 0.60\\
\hline
800 & 0.00 & 109993 & 0.24 & 0.00 & 37660 & 0.24 & 0.00 & 4444 & 0.24 & 0.00\\
800 & 0.20 & 109598 & 0.24 & 0.00 & 20891 & 0.25 & 0.08 & 4490 & 0.36 & 0.37\\
800 & 0.40 & 106535 & 0.24 & 0.00 & 20724 & 0.31 & 0.24 & 4525 & 0.40 & 0.46\\
800 & 0.60 & 101675 & 0.24 & 0.00 & 36104 & 0.36 & 0.36 & 4559 & 0.45 & 0.57\\
\hline
\end{longtable}

\begin{longtable}{c c c c c c c c c c}
 \caption{Statistics of clusters averaged across 10,000 trials. $n$: the cluster multiplicity; $M_{\mathrm{total}}$: the combined mass of stars in the cluster; $m_{\mathrm{z}}$: the magnitude at redshift $z$ in the best band; $E_{\mathrm{z}}$: the color at redshift $z$ in the best pair of bands}.\\
 \hline
$n$ & $M_{\mathrm{total}}$ & $m_\mathrm{z=3}$ & $E_\mathrm{z=3}$ & $m_\mathrm{z=7}$ & $E_\mathrm{z=7}$ & $m_\mathrm{z=12}$ & $E_\mathrm{z=12}$ & $m_\mathrm{z=17}$ & $E_\mathrm{z=17}$ \\ & ($M_\odot$) & & & & & & & & \\ \hline\endfirsthead
 \caption{(Continued)}\\
 \hline
$n$ & $M_{\mathrm{total}}$ & $m_\mathrm{z=3}$ & $m_\mathrm{z=7}$ & $m_\mathrm{z=12}$ & $m_\mathrm{z=17}$ & $E_\mathrm{z=3}$ & $E_\mathrm{z=7}$ & $E_\mathrm{z=12}$ & $E_\mathrm{z=17}$ \\ & ($M_\odot$) & & & & & & & & \\
 \hline
 \endhead
 \label{tab:cluster}
5 & $128 \pm 63$ & $39.7 \pm $0.9 & $-0.408 \pm $0.022 & $40.8 \pm $0.9 & $-0.418 \pm $0.019 & $41.7 \pm $0.9 & $-0.763 \pm $0.034 & $42.2 \pm $0.9 & $-0.405 \pm $0.019\\
10 & $255 \pm 90$ & $38.8 \pm $0.6 & $-0.414 \pm $0.012 & $39.9 \pm $0.6 & $-0.423 \pm $0.009 & $40.8 \pm $0.6 & $-0.772 \pm $0.018 & $41.3 \pm $0.6 & $-0.409 \pm $0.010\\
20 & $511 \pm 128$ & $37.9 \pm $0.4 & $-0.417 \pm $0.007 & $39.0 \pm $0.4 & $-0.424 \pm $0.005 & $39.9 \pm $0.4 & $-0.775 \pm $0.009 & $40.4 \pm $0.4 & $-0.411 \pm $0.005\\
30 & $764 \pm 158$ & $37.5 \pm $0.3 & $-0.417 \pm $0.005 & $38.6 \pm $0.3 & $-0.425 \pm $0.003 & $39.5 \pm $0.3 & $-0.776 \pm $0.007 & $40.0 \pm $0.3 & $-0.412 \pm $0.004\\
40 & $1016 \pm 183$ & $37.2 \pm $0.3 & $-0.418 \pm $0.004 & $38.3 \pm $0.3 & $-0.425 \pm $0.003 & $39.2 \pm $0.3 & $-0.777 \pm $0.006 & $39.7 \pm $0.3 & $-0.412 \pm $0.003\\
50 & $1270 \pm 201$ & $36.9 \pm $0.3 & $-0.418 \pm $0.003 & $38.0 \pm $0.3 & $-0.425 \pm $0.002 & $38.9 \pm $0.3 & $-0.777 \pm $0.005 & $39.4 \pm $0.3 & $-0.412 \pm $0.003\\
60 & $1525 \pm 221$ & $36.7 \pm $0.2 & $-0.418 \pm $0.003 & $37.8 \pm $0.2 & $-0.425 \pm $0.002 & $38.7 \pm $0.2 & $-0.777 \pm $0.004 & $39.2 \pm $0.2 & $-0.412 \pm $0.002\\
70 & $1780 \pm 240$ & $36.5 \pm $0.2 & $-0.418 \pm $0.003 & $37.6 \pm $0.2 & $-0.425 \pm $0.002 & $38.5 \pm $0.2 & $-0.777 \pm $0.004 & $39.0 \pm $0.2 & $-0.412 \pm $0.002\\
80 & $2037 \pm 255$ & $36.4 \pm $0.2 & $-0.418 \pm $0.003 & $37.5 \pm $0.2 & $-0.425 \pm $0.002 & $38.4 \pm $0.2 & $-0.777 \pm $0.004 & $38.9 \pm $0.2 & $-0.412 \pm $0.002\\
\hline
\end{longtable}

\section{Massive stars near the end of main sequence}\label{appendix:tams}
As discussed in Section \ref{sec:methods}, massive stars on the RSG track may quickly move redward near the end of the main sequence, due to developing large convective cores. Because many of these stars are too massive to achieve CHE at any rotational velocity, this means very massive stars will generally exhibit this redward shift even at high rotational velocities. As a result, redder bands than those listed in Table \ref{tab:limitingmags} are needed to optimally detect many massive stars near TAMS. In this section, we consider the observability of these models in their optimal bands.
\begin{table}
\begin{center}
    \begin{tabular}{|c |c c|c c | } 
 \hline
 $z$ & Band 1 & $m_\mathrm{lim}$ & Band 2 & $m_\mathrm{lim}$ \\
 \hline
 3 & NIRCAM F277W & 30.43 & NIRCAM F356W & 30.47 \\ 
 7 & NIRCAM F360M & 30.05 & NIRCAM F444W & 29.70 \\ 
 12 & MIRI F770W & 27.08 & MIRI F1000W & 26.51 \\ 
 17 & MIRI F1000W & 26.51 & MIRI F1280W & 25.94 \\ 
 \hline
\end{tabular}
\caption{Limiting magnitudes for the optimal pairs of JWST bands for high-mass stars near TAMS, at each redshift.}
\label{tab:limitingmags2}
\end{center}
\end{table}

We find that the pairs of bands used in Figure \ref{fig:colormagnitude} are optimal for all of our ZAMS models and nearly all of our IAMS models. For $\Omega=0.0, 0.2$, they are still the optimal pairs for TAMS models with masses up to $M_{\mathrm{init}} \lesssim 100\;\mathrm{M}_\odot$; for $\Omega=0.4$, they remain optimal for models up to $M_{\mathrm{init}} \lesssim 150\; \mathrm{M}_\odot$; and for $\Omega=0.6$, they remain optimal up to $M_{\mathrm{init}} \lesssim 350 \;\mathrm{M}_\odot$. At masses above these thresholds, a second set of bands becomes optimal, which generally depends on the redshift alone. We provide a list of them in Table \ref{tab:limitingmags2}.
\begin{figure*}[ht!]
\centering
\includegraphics[width=\textwidth]{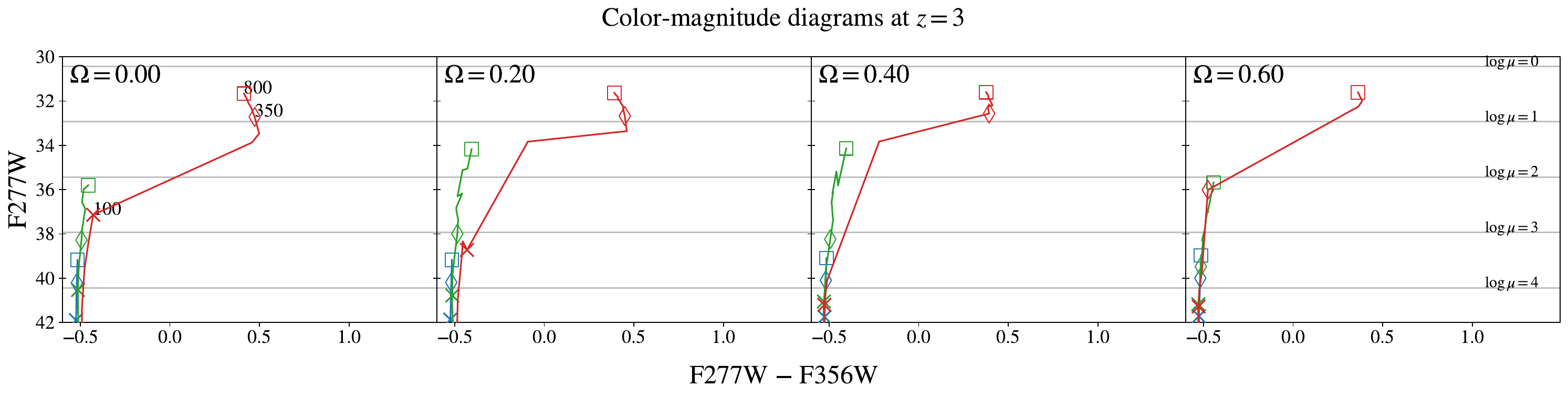}
\includegraphics[width=\textwidth]{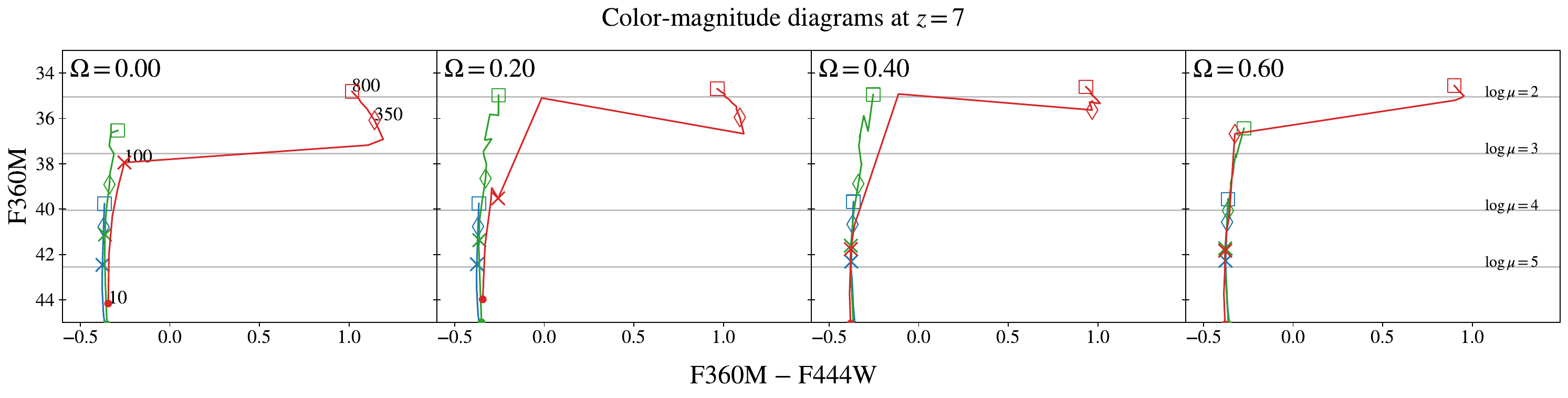}
\caption{Color-magnitude diagrams in the optimal bands for massive stars near TAMS, at redshifts $z=3$ and $z=7$.}
\label{fig:colormagnitude_hi}
\end{figure*}

Figure \ref{fig:colormagnitude_hi} shows color-magnitude diagrams of IAMS and TAMS models in the bands from Table \ref{tab:limitingmags2}, at $z=3$ and $z=7$. For $z=3$, the high-mass TAMS models are much easier to detect in their optimal bands compared to the corresponding ZAMS and IAMS models shown in Figure \ref{fig:colormagnitude}. Several of these models are detectable even at lensing amplifications below $\mu \sim 10$. At $z=7$, many of these models are still easier to detect than their ZAMS and IAMS counterparts in Figure \ref{fig:colormagnitude}; however, the most massive models in our study do not see a significant improvement, and the lower bound on $\mu$ needed to detect Population III stars is not substantially improved.

At $z=12$ and $z=17$, TAMS models are in fact less detectable with JWST than their ZAMS and IAMS counterparts. At these redshifts, the magnitudes of the TAMS models peak in redder bands than those listed in Table \ref{tab:limitingmags2}. However, the limiting magnitudes for MIRI bands quickly fall as the wavelength increases, making the peak bands unsuitable for detecting our models.

Ultimately, we find that using the pairs from Table \ref{tab:limitingmags2} only improves the detectability for a subset of Population III stars satisfying all of the following requirements:

\noindent
{\em (i)} The stars are at lower redshifts ($z\lesssim7$);\\
{\em (ii)} The stars are very massive (especially if they have high rotational velocities);\\
{\em (iii)} The stars have exhausted the majority of the hydrogen in their cores.\\
Because of these restrictions, we expect that such cases are rare, and thus that the bands in Table \ref{tab:limitingmags} are generally better suited for detecting Population III stars with JWST. Beyond JWST, instrumentation more suited to mid- or far-infrared wavelengths may be used to detect near-TAMS massive stars at higher redshifts. Additionally, while this study focused on main-sequence stars, we note that detecting red supergiants may require similar instrumentation, due to the proximity of TAMS with post-MS phases for massive stars on the H-R diagram.

\bibliography{biblio}

\end{document}